\documentclass[preprint,12pt]{elsarticle}

\def\mZ{\mathbb{Z}}

\def\be{\begin{eqnarray}}
\def\ee{\end{eqnarray}}

\def\wH{\widetilde{H}}

\usepackage{amssymb}

\journal{Physics Letters B}

\begin{document}

\begin{frontmatter}



\title{Pairs of chiral quarks on the lattice from staggered fermions}


\author{David H.~Adams}

\address{Division of Mathematical Sciences, Nanyang Technological
University, Singapore 637371}

\begin{abstract}
A new formulation of chiral fermions on the lattice is presented. It is a 
version of overlap fermions, but built from the computationally efficient 
staggered fermions rather than the previously used Wilson fermions.
The construction reduces the four quark flavors described by the 
staggered fermion to two quark flavors; 
this pair can be taken as the up and down quarks in Lattice QCD.
A domain wall formulation giving a truncation of this overlap 
construction is also outlined. 

\end{abstract}

\begin{keyword}
Lattice QCD \sep chiral fermions \sep staggered fermions


\end{keyword}

\end{frontmatter}

\section{Introduction}

It is of great interest to find lattice formulations of Quantum Chromodynamics 
(QCD) with exact chiral symmetry. In such a formulation, 
spontaneous chiral symmetry breaking and the associated spectrum of 
Goldstone bosons (i.e. the light mesons), as well as the emergence of the 
flavor-singlet ($\eta'$ meson) mass via the index theorem connection
between quark zero-modes and gauge field topology \cite{WV},  
can be studied directly in the lattice 
model without the usual need to extrapolate via chiral perturbation theory.
Finding such formulations is problematic though \cite{NN}, and was a 
long-standing challenge in Lattice QCD. An explicit formulation of this kind
was finally found by Neuberger \cite{Neu(PLB)} via the overlap formulation
of chiral fermions on the lattice \cite{overlap}. 
It is a chirally improved version of Wilson fermions, and realizes exact chiral
symmetry of the Ginsparg-Wilson (GW) type \cite{GW,Has(GW),Luscher(PLB)}. 
However, its theoretical attractiveness is offset in practice by the 
high computational cost of implementing it in simulations.

In this paper, building on insights from \cite{DA(PRL)}, 
a new version of overlap fermions is presented constructed from the 
staggered lattice fermion rather than the previously used Wilson fermion. 
The staggered fermion is the most computationally efficient lattice fermion
formulation known, and this advantage is expected to be inherited 
by the new staggered version of overlap fermions. 
The challenge in doing this is that the overlap construction relies on 
specific properties of Wilson fermions that are not shared by staggered
fermions. This challenge is overcome here using a new theoretical
idea of interest in its own right: a Wilson-like
phase of the staggered fermion theory can be created by adding a certain
``flavored'' mass term. In this new phase, the number of quark flavors 
described by the staggered fermion is reduced from four to two, and the 
staggered fermion acquires Wilson-like properties. The latter allow the overlap
construction to work as desired. The resulting staggered overlap fermion
describes two quark flavors and has an exact {\em unflavored} GW chiral
symmetry originating from the exact flavored chiral symmetry of the
original staggered fermion.

However, the {\em flavored} vector and chiral symmetries of the 
quark pair described by the staggered overlap fermion are not exact since 
the corresponding symmetries of the original staggered fermion are broken by 
lattice effects. Therefore, exact flavored vector and chiral symmetries only 
hold for {\em pairs} of quark flavors when each pair is described by a 
staggered overlap fermion.

The two quark flavors described by the staggered overlap fermion can be taken
as the almost massless $u$ and $d$ quarks in QCD. Then, due to the exact
unflavored GW chiral symmetry, this description has all the advantages
of the usual overlap fermions regarding the axial U(1) anomaly and index 
theorem connection between quark zero-modes and gauge field topology. This is 
connected with some of the most subtle and interesting 
parts of QCD physics -- not only the $\eta'$ mass mentioned already but 
also the order of the QCD phase transition at finite temperature -- which 
are also computationally highly demanding 
to study in lattice QCD (see, e.g., \cite{Christ(temp)}).
Finding a new lattice approach in which these features are not distorted by
lattice effects, and which is computationally more efficient than usual
overlap fermions, is therefore significant. As an indication of the 
desirability of this, it can be noted that high-precision calculation of
the $\eta'$ mass is still an outstanding problem in Lattice QCD --
the latest state of the art 
calculation, using Wilson-based domain wall fermions
(a truncated version of usual overlap fermions), has an uncertainly of 15\% 
\cite{Christ(eta)}. 

 

\section{Staggered overlap fermion construction}

The staggered lattice fermion field is a one component Grassmann field
$\chi(x)$, $x=an\in a\mZ^4$, describing
4 degenerate Dirac fermions. The staggered Dirac operator acting on 
these fields is $D_{st}=\eta_{\mu}\nabla_{\mu}$
where $\eta_{\mu}\chi(x)=(-1)^{n_1+\dots+n_{\mu-1}}\chi(x)$ and 
$\nabla_{\mu}=\frac{1}{2a}(T_{+\mu}-T_{-\mu})$, with the parallel transporters
$T_{\pm\mu}$ given in terms of the lattice link variables $U_{\mu}(x)$ by 
$T_{+\mu}\chi(x)=U_{\mu}(x)\chi(x+a\hat{\mu})$ and $T_{-\mu}=(T_{+\mu})^{-1}$.
There is an exact {\em flavored} chiral symmetry 
\be
\{\Gamma_{55}\,,D_{st}\}=0
\label{3}
\ee
where $\Gamma_{55}$, given by
$\Gamma_{55}\chi(x)=(-1)^{n_1+\dots+n_4}\chi(x)$,
corresponds to $\gamma_5\otimes\gamma_5$ in the spin-flavor interpretation
of staggered fermions \cite{GS}. The first and second $\gamma_5$
factors act in spinor and flavor space, respectively. Note
the properties $\Gamma_{55}^2={\bf 1}$ and $\Gamma_{55}^{\dagger}=\Gamma_{55}$.

Direct application of the overlap construction \cite{Neu(PLB)} to staggered 
fermions by simply replacing the kernel operator $D_W-M$ (where $D_W$ is
the Wilson-Dirac operator) by $D_{st}-M$ does not 
give anything useful: The exact flavored chiral symmetry of the staggered 
fermion is lost, there is no GW symmetry to replace it, and there are no exact 
zero-modes in general. However, as we will now show, the situation changes
if, instead of a scalar mass $M$, we use the following 
``flavor-chiral'' mass term:
\be
M_{st}=\frac{r}{a}\Gamma_{55}\Gamma_5\quad,\quad r>0
\label{5}
\ee 
Here $\Gamma_5$
is the staggered fermion version of the chirality matrix, corresponding
to $\gamma_5\otimes{\bf 1}$ up to $O(a^2)$ discretization errors in the
spin-flavor interpretation \cite{GS}. 
Consequently, in the spin-flavor interpretation, 
\be
M_{st}\ \sim\ {\bf 1}\otimes\gamma_5+O(a).
\label{8}
\ee
Explicitly, $\Gamma_5$ is given as follows. 
Use the parallel transporters to define 
$C_{\mu}=\frac{1}{2}(T_{+\mu}+T_{-\mu})$, and let
$C=(C_1C_2C_3C_4)_{sym}$ denote the symmetrized product of $C_1,C_2,C_3,C_4$.
Then $\Gamma_5=\eta_5C$ where $\eta_5=\eta_1\eta_2\eta_3\eta_4$.
(Explicitly, $\eta_5\chi(x)=(-1)^{n_1+n_3}\chi(x)$.)
Note that $\Gamma_5$ is hermitian and commutes with 
$\Gamma_{55}$; therefore $M_{st}$ is hermitian.
The effect of this mass term in 
\be
D_{st}-M_{st}
\label{9}
\ee
is to split the low-lying modes into branches with approximately definite
positive and negative flavor-chirality, giving them masses $-\frac{r}{a}$
and $\frac{r}{a}$, respectively. This is shown in the free field case in
Fig.~1, taken from \cite{Forcrand}. 
\begin{figure}
\begin{center}
\includegraphics[width=1.3in]{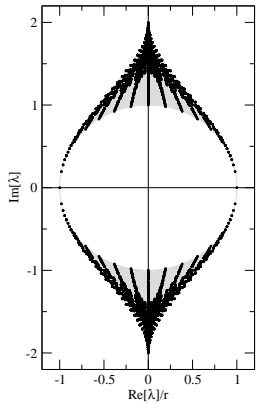}
\caption{\label{forcrandeps} Spectrum of $D_{st}-M_{st}$ in the free field 
case \cite{Forcrand}.}
\end{center}
\end{figure} 
In this case, there are 16 zero-modes of 
$D_{st}$ with momenta $p_A=\frac{\pi}{a}A\,$, $A_{\mu}\in\{0,1\}$.
On the vectorspace $V$ spanned by these modes we have 
$(\Gamma_{55}\Gamma_5)^2={\bf 1}$, so there is a decomposition 
$V=V_+\oplus V_-$ with $\Gamma_{55}\Gamma_5=\pm{\bf 1}$ on $V_{\pm}$,
and it can be shown that $\dim(V_{\pm})=8$.  
Hence $D_{st}-M_{st}=\mp\frac{r}{a}$ on $V_{\pm}$; this shows how the real 
eigenvalues arise in the free field case. 
Moreover, the operators $\eta_{\mu}T_{+\mu}$ and  $\eta_{\mu}T_{-\mu}$
making up $D_{st}$ coincide on $V$ and can be shown to commute with 
$\Gamma_{55}\Gamma_5$ on $V$. 
Hence their representations $\Gamma_{\mu}$ on $V$ (given in \cite{GS} by the 
$16\times16$ matrices $(\Gamma_{\mu})_{AB}$), which form a representation of 
the Euclidean Dirac algebra \cite{GS}, decompose into representations
$\Gamma_{\mu}^{\pm}$ on $V_{\pm}$. 
From this it can be seen that, from a low momentum viewpoint, the 4 Dirac 
fermion species described by the staggered fermion are split into two
pairs of Dirac fermions, with the pair with positive (negative) 
flavor-chirality having mass $-\frac{r}{a}$ ($+\frac{r}{a}$).

Upon inserting the kernel $D_{st}-M_{st}$ into the overlap formula, 
\be
D_{so}&=&\frac{r}{a}\Big(1+(D_{st}\!-\!M_{st})\frac{1}
{\sqrt{(D_{st}\!-\!M_{st})^{\dagger}(D_{st}\!-\!M_{st})}}\Big),
\nonumber \\
&&\label{10}
\ee
the negative mass modes are converted into the physical modes of a massless
fermion, while the positive mass modes are converted into modes with heavy 
masses $\sim 1/a$ which decouple in the continuum limit, just as in the Wilson 
case. Thus only two of the original four quark flavors survive in the resulting
overlap fermion. Furthermore, since the physical modes have approximately 
definite positive flavor-chirality, $\Gamma_{5}=\Gamma_{55}$ on these modes 
up to $O(a^2)$ effects. 

The staggered overlap Dirac operator (\ref{10}) satisfies the following
GW relation:
\be
\{\Gamma_{55}\,,D_{so}\}=\frac{a}{r}D_{so}\Gamma_{55}D_{so}
\label{11}
\ee
Note that this is now an {\em unflavored} chiral symmetry, since, as remarked, 
$\Gamma_{55}=\Gamma_5+O(a^2)$ on the physical modes of the staggered overlap
fermion. The GW relation can be readily verified
after noting that the staggered overlap operator can be expressed as
\be
D_{so}&=&\frac{r}{a}\Big(1+\Gamma_{55}\frac{H_{st}}
{\sqrt{H_{st}^2}}\Big)
\label{12}
\ee
with the hermitian operator $H_{st}$ given by
\be
H_{st}=\Gamma_{55}(D_{st}-M_{st})=\Gamma_{55}D_{st}-\frac{r}{a}\Gamma_5
\label{13}
\ee
It also follows from (\ref{12}) that 
$D_{so}^{\dagger}=\Gamma_{55}D_{so}\Gamma_{55}$, which implies that the 
non-real eigenvalues of $D_{so}$ come in complex conjugate pairs.
Since the only possible real eigenvalues (which must lie on the GW circle) are 
$0$ and $2r/a$, it follows that the fermion determinant $\det(D_{so})$ is 
real and non-negative, as required for lattice QCD simulations.


A required property of the lattice Dirac operator is locality. It holds
for $D_{so}$; this can be proved analytically by the same argument as in 
the Wilson overlap case \cite{L(local)}, using the staggered bound mentioned 
in \cite{DA(PRL)}, when the lattice gauge fields satisfy an admissibility 
condition; the details are in \cite{DA(soverlap)}. Locality of $D_{so}$ in
realistic gauge field backgrounds has been shown numerically in
\cite{Forcrand}.

Computational efficiency of $D_{so}$ was also studied in \cite{Forcrand}:
preliminary numerical evidence indicates a speedup of order 2-3 compared to 
the Wilson-based overlap Dirac operator in realistic gauge field backgrounds.

\section{Pairs of exact zero-modes and index theorem}

We now verify that $D_{so}$ has pairs of exact chiral zero modes with 
index determined by gauge field topology in accordance with the 2-flavor 
index theorem when the lattice gauge field background is not too rough.
Due to the GW relation (\ref{11}), the index is well-defined as
$\;\mbox{index}(D_{so})=n_+-n_-\;$
where $n_{\pm}$ is the number of independent zero-modes of $D_{so}$ with 
$\pm$ chirality under $\Gamma_{55}$ (which is the same as $\Gamma_5$ on these
modes up to $O(a^2)$ effects). A general formula \cite{Has(GW)} gives
\be
\mbox{index}(D_{so})=-\frac{1}{2}\mbox{Tr}\Big(
\frac{H_{st}}{\sqrt{H_{st}^2}}\Big).
\label{16}
\ee
We connect this to the hermitian staggered operator
studied previously in \cite{DA(PRL)}, denoted by $\wH_{st}$ below:

\noindent {\bf Theorem.} {\em For every value of $m$, the hermitian staggered 
operators}
\be
H_{st}(m)=\Gamma_{55}D_{st}-m\Gamma_5\ ,\ \,   
\wH_{st}(m)=iD_{st}-m\Gamma_5 
\nonumber
\ee
{\em have the same eigenvalue spectrum.} 

The proof of the theorem is deferred to \cite{DA(soverlap)}. It can be 
understood intuitively from the fact that $\wH_{st}(m)$ arises from 
$H_{st}(m)$ by a change of representation of the staggered sign factors
$\eta_{\mu}\to i\Gamma_{55}\eta_{\mu}$ in $D_{st}$.

In light of the theorem we can replace $H_{st}$ in (\ref{16}) by 
$\wH_{st}(m)$ with $m=\frac{r}{a}$; here $r$ is the parameter in (\ref{5}). 
But then the results of \cite{DA(PRL)} give
\be
\mbox{index}(D_{so})={\textstyle \frac{1}{2}}\mbox{index}(D_{st})=2Q
\label{18}
\ee
in sufficiently smooth gauge field backgrounds of topological 
charge $Q$ and with $\frac{r}{a}$ lying in a suitable range. 
The last equality in (\ref{18}) was confirmed in the numerical study 
in \cite{DA(PRL)}.

\section{Stability of the massless phase under radiative corrections}

It is important to consider the symmetries of the staggered overlap fermion
since these need to be sufficient for renormalizability and stability of
the massless 2-flavor phase under radiative corrections.
Besides the replacement of the flavored chiral symmetry (\ref{3}) by the
GW symmetry (\ref{11}), the usual staggered fermion symmetries listed in
\cite{GS} all hold for the staggered overlap fermion, except for axis
reversals and shift transformations. Under the latter 
$\bar{\chi}\Gamma_{55}\Gamma_5\chi\to-\bar{\chi}\Gamma_{55}\Gamma_5\chi$,
and therefore $\bar{\chi}D_{so}\chi$ is invariant when these transformations 
are combined with the parameter flip $r\to-r$ in (\ref{5}).
The same invariances must hold for all counterterms that arise. Hence the
counterterms must be invariant under all the usual staggered fermion 
symmetries except (\ref{3}), and except for a possible sign change under
axis reversals and shift transformations. It is easy to show 
\cite{DA(soverlap)} that the most general {\em local} mass-dimension 4 terms
with these properties are (up to proportionality, and modulo $O(a)$ 
terms)\footnote{This is with the lattice gauge fields included. In the free 
field case $\Gamma_{55}\Gamma_5$ and $D_{st}$ commute and the second term in 
(\ref{18a}) simplifies.}
\be
\bar{\chi}D_{st}\chi\ ,\ \ \bar{\chi}\{\Gamma_{55}\Gamma_5,D_{st}\}\chi\ ,\ \
\frac{1}{a}\bar{\chi}\Gamma_{55}\Gamma_5\chi\ ,\ \ \frac{1}{a}\bar{\chi}\chi
\label{18a} 
\ee
Since $D_{so}$ is local, all counterterms are also local, and the theory is
renormalizable, provided the power-counting theorem holds \cite{Reisz1}. 
It holds in the Wilson overlap case \cite{Reisz2} and is plausible in the 
present case. Assuming this, stability can be established to all 
orders in the loop expansion as follows. 
The GW relation (\ref{11}) leads to
\be
\Gamma_{55}\langle D_{so}^{-1}(q)\rangle
+\langle D_{so}^{-1}(q)\rangle\Gamma_{55}
=\frac{a}{r}\Gamma_{55}
\label{18b}
\ee
where $\langle D_{so}^{-1}(q)\rangle=(D_{so}(q)+\Sigma(q))^{-1}$ with 
$\Sigma(q)$ being the fermion self-energy. Here and
in the following, $X(q)$ denotes the free field momentum representation of
the staggered operator $X$; it is a linear map ($16\times16$ matrix) 
$X(q):V\to V$ on the zero-mode vector space $V$ defined earlier, with 
$q_{\mu}\in[-\frac{\pi}{2a},\frac{\pi}{2a}]$.
The momentum representation of $\Gamma_{55}$ is a constant matrix, and 
$\Gamma_{55}\Gamma_5(q)=\pm C(aq){\bf 1}$ on $V_{\pm}$ where 
$C(aq)=\prod_{\mu}\cos(aq_{\mu})=1+O(a^2q^2)$.

In light of the preceding, at one loop $\Sigma(q)$ has the form
\be
c_1D_{st}(q)+c_2\Gamma_{55}\Gamma_5(q)D_{st}(q)
+\frac{c_3}{a}\Gamma_{55}\Gamma_5(q)+\frac{c_4}{a}
\label{18c}
\ee
up to irrelevant terms, with each $c_j=c_j(a^2q^2)$ diverging no worse than 
logarithmically for $a\to0$. Note that all terms in (\ref{18c}) map 
$V_{\pm}$ to itself, as does $D_{so}(q)$.
From (\ref{18b})--(\ref{18c}) and the fact that
$D_{so}(0)=0$ on $V_+$ it is straightforward to show $\Sigma(0)=0$ on $V_+$. 
(We omit the details; an analogous result was found in the Wilson overlap 
case in \cite{Reisz2,Chiu-KY}). This implies $c_4=-c_3$ in (\ref{18c}), so 
on $V_+$ the last two terms in (\ref{18c}) combine to give an irrelevant 
term $\frac{c_3}{a}(C(aq)-1)$ which vanishes $\sim a\log(a)$ for $a\to0$.
The first two terms in (\ref{18c}) act on $V_{\pm}$ as 
$(c_1\pm c_2C(aq))(\Gamma_{\mu}^{\pm}\frac{i}{a}\sin(aq_{\mu}))$.
It follows that, for $a\to0$,
$D_{so}(q)+\Sigma(q)=\Gamma_{\mu}^+iq_{\mu}(1+c_1(a^2q^2)+c_2(a^2q^2))$
on $V_+$. The self-energy is hereby seen to have 
the usual effect of a logarithmically divergent wave function renormalization 
on the massless physical modes. Then, after renormalizing the theory at one 
loop, the argument above can be repeated to get the same result at two
loops, and so on. In this way stability is seen to hold to all orders
in the renormalized theory.

\section{Wilson-like nature of the staggered overlap fermion kernel}

A deeper understanding of the staggered overlap construction can be obtained
by clarifying the Wilson-like nature of its staggered fermion kernel. 
To this end we introduce the following 2-flavor Wilson-like version of the 
staggered Dirac operator:
\be
D_{sW}=D_{st}+W_{st}\quad,\quad W_{st}=\frac{r}{a}(1-\Gamma_{55}\Gamma_5).
\label{19}
\ee
The term $W_{st}$ plays an analogous role to the Wilson term in the 
Wilson-Dirac operator: it decouples the negative
flavor-chirality modes by giving them mass $2r/a$ while keeping the
two positive flavor-chirality modes as the physical modes.
Hence $D_{sW}$ describes two physical quark flavors on which 
$\Gamma_5=\Gamma_{55}$ up to $O(a)$ effects. 
In light of this we can obtain a 2-flavor overlap fermion 
by taking $D_{sW}-M$ with $M=r\rho/a$, $\rho\in(0,2)$
as the kernel in the usual overlap construction.  
For $\rho=1$ this is precisely the 2-flavor staggered overlap Dirac operator
$D_{so}$ we constructed above in (\ref{10}), since
$
D_{sW}-\frac{r}{a}=D_{st}-M_{st}.
$
But now we see that it can be
generalized to any $\rho\in(0,2)$. Furthermore, the role of the parameter
$r$ in the staggered overlap construction is hereby clarified: it is 
analogous to the Wilson parameter in the usual overlap construction based 
on Wilson fermions. This is surprising, since $r/a$ in (\ref{5}) 
initially appears to be analogous to the mass parameter $M=\rho/a$
in the Wilson case. 

Starting from Wilson fermions, a domain wall fermion formulation can be
constructed \cite{Shamir} which gives a truncation of the overlap fermion
construction \cite{Neu(vector)}. The same can now be done with staggered
fermions simply by replacing
\be
D_W\;\to\;D_{sW}\quad,\quad \gamma_5\;\to\;\Gamma_{55}
\label{26}
\ee
in the previous Wilson-based constructions.
The lattice Dirac operator for the staggered domain wall fermion
in 5 dimensions is then 
\be
D_{sdw}=D_{sW}-M+\Gamma_{55}\tilde{\partial}_s\ ,\quad M=\frac{r\rho}{a}
\ ,\ \rho\in(0,2)
\label{27}
\ee
where $s\in[0,L]$ is the lattice coordinate of the 5th dimension and
$\tilde{\partial}_s=P_+\partial_s^{(+)}+P_-\partial_s^{(-)}$ with 
$P_{\pm}=\frac{1}{2}(1\pm\Gamma_{55})$ and $\partial_s^{(+)}\,$, 
$\partial_s^{(-)}$ being the forward and backward finite difference 
operators.\footnote{This is completely unrelated to a previous proposal for 
staggered domain wall fermions in \cite{Fleming}  
which describes 4 rather than 2 flavors.}
Boundary conditions on $\chi(x,s)$ at $s=0$ and $s=L$ are introduced 
analogously to the Wilson-based case; then the staggered domain wall
construction is seen to be a truncation of the staggered overlap construction
by essentially the same argument as in \cite{Neu(vector)}.
The details will be given in \cite{DA(soverlap)}.

\section{Concluding discussion}

Underlying the staggered overlap and staggered domain wall fermion
constructions in this paper is a staggered version of Wilson fermions
obtained by adding a flavored mass term to the staggered fermion action.
The idea of considering staggered fermion with flavored mass term is
not new in itself; this was already done many years ago by Golterman and
Smit in \cite{GS}. Rather, the new idea here is that, for a certain choice
of flavored mass term, namely the Wilson term $W_{st}$ in (\ref{19}) 
determined by the ``flavor-chiral'' mass term $M_{st}$ in (\ref{5}),
one can use the staggered operator $\Gamma_{55}$ for the role of $\gamma_5$
since it coincides with $\gamma_5\otimes{\bf 1}$ on the 2 physical flavors
up to $O(a)$ effects.
The significance of this is that the properties $\gamma_5^2={\bf 1}$ and 
$\gamma_5-$hermiticity, which are crucial in the usual overlap and domain
wall constructions, continue to hold in the staggered setting with 
$\gamma_5\to\Gamma_{55}$. This would not be true if one tried to use
the direct staggered analogue $\Gamma_5$ of $\gamma_5$.    

The flavor-chiral mass term used in this work, which reduces the 4 staggered
fermion flavors to 2 flavors, is not the only possibility. 
Another flavor-chiral mass term which reduces the number of flavors to 1 
was subsequently proposed in \cite{Hoelbling} after the first version
of this paper appeared on the Arxiv. However, it breaks more of the staggered
fermion symmetries than the present proposal, namely some of the lattice 
rotation symmetries. The consequences of this for radiatively generated 
counter-terms and fine-tuning requirements remains to be seen. 
The 1-flavor version of staggered overlap fermions has the advantage that 
all the {\em flavored} vector and GW-chiral symmetries are exact.
On the other hand, for problems where only the {\em unflavored} chiral
symmetry is important, e.g. for computing the $\eta'$ mass, the present
2-flavor formulation has advantages: it preserves lattice
rotation symmetry and gives ``two quarks (the $u$ and $d$) for the 
price of one'' in lattice QCD simulations.

\medskip

\noindent {\bf Acknowledgments}

\medskip

I thank Mike Creutz, Philippe de Forcrand, Maarten Golterman and 
Herbert Neuberger for discussions and/or feedback.
I also thank Philippe de Forcrand and his group for performing the numerical 
investigations in \cite{Forcrand} and for kindly providing me with Fig.~1.
This work was supported by the AcRF Tier 1 grant RG 61/10.
Some of this work was done at the National Center for Theoretical Sciences, 
Taiwan.

\end{document}